\documentstyle[prl,aps,floats]{revtex}
\input epsf

\begin{document}

% \draft command makes pacs numbers print
\draft
\newcommand{\vertsp}{\vphantom{\displaystyle{\dot a \over a}}}
\newcommand{\Spin}[4]{\, {}_{#2}^{\vphantom{#4}} {#1}_{#3}^{#4}}

\title{Cosmic defects and CMB anisotropy}

% repeat the \author\address pair as needed

\author{Levon Pogosian}
\address{
Physics Department, 
Case Western Reserve University, 
Cleveland OH 44106-7079.
}

%\date{\today}

%\twocolumn[
\wideabs{

\maketitle

%\tightenlines

\begin{abstract}
\widetext
Recent measurements of the cosmic microwave background (CMB) anisotropies by 
BOOMERANG and MAXIMA collaborations have tightened the observational
constraints on theories of structure formation. They disagree with the 
predictions of conventional topological defect models.
Considering the fact that topological defects are predicted by the majority 
of realistic particle physics models, the exact nature of the constraints 
imposed by the recent data on the population and the properties of the defects 
must be fully understood.
We show that the predictions of current cosmic string models can be brought 
into a closer agreement with the observations by choosing a 
closed universe with $\Omega=1.3$ and by including the effects of the 
small-scale structure and radiation products of the strings. These alone, 
however, are not sufficient for obtaining a good fit to the measured shape 
of the angular power spectrum. To fit the data cosmic strings would either have
to be correlated on large (perhaps superhorizon) scales or would have to
possess a higher degree of coherence, {\it i.e.} be more ``time-correlated''.

\end{abstract}

%insert suggested PACS numbers in braces on next line
\pacs{}

%]
}

\narrowtext

\section{Introduction}
\label{introduction}

Observations of the Cosmic Microwave Background (CMB) Anisotropy have catalyzed
the rapid growth in the field of Cosmology over the last
decade. They have given theorists a rare opportunity to test their 
models against experiment. In particular, the CMB data can help identify 
the mechanism that led to the formation of large scale
structure that we observe today. 

The recent measurements of the CMB anisotropy by BOOMERANG \cite{boomerang} 
and MAXIMA \cite{maxima} experiments have set a new precision standard in 
comparing the predictions of different theories.
There are two main classes of models of structure formation - the inflationary 
models and models with topological defects.
The experimental data has different implications for this two classes of 
models. If one assumes an inflationary scenario then the data can be fit 
quite well by choosing adiabatic initial conditions, a nearly flat universe 
with a cosmological constant and a relatively high baryon density \cite{jaffe}. 
Our goal is to understand the implications of the data for defect models.

It is widely believed that at sufficiently high energies all known forces
of nature must be described by a single Grand Unified Theory (GUT) of
fundamental interactions. It is also believed that the Universe evolved
from a very hot state (with temperatures well above the GUT scale) and 
underwent a series of phase transitions as it cooled down due to the expansion.
Topological defects would inevitably form during these phase transitions, 
hence their subsequent evolution and observational signatures must be 
understood.
Defects provide the only alternative to inflation as a mechanism for
generating density fluctuations. Despite its popularity, inflation still 
remains to be placed in a realistic particle physics context. Inflationary 
models can be carefully tailored to fit practically any observational data 
by a suitable choice of otherwise arbitrary parameters. In contrast, 
the defect models have the attractive feature that they have no 
parameter freedom as all the necessary information is in principle 
contained in the underlying particle physics model.

Calculating CMB anisotropies sourced by topological defects is a rather
difficult task. In inflationary scenario the entire 
information about the seeds is contained in the initial conditions for the 
perturbations in the metric.
In the case of cosmic defects, perturbations are continuously seeded over 
the period of time from the phase transition that 
had produced them until today. The exact determination of 
the resulting anisotropy requires, in principle, the knowledge of the 
energy-momentum tensor (or, if only two point functions are being calculated,
the unequal time correlators \cite{uetc}) of the defect network and 
the products of its decay at all times. This information is simply not 
available! Instead, a number of clever simplifications, based on the expected 
properties of the defect networks ({\it e.g.} scaling), are used to 
calculate the source. 
The latest data from BOOMERANG and MAXIMA experiments clearly disagree
with the predictions of these simple models of defects.

The approach that we would like to take is not to simply 
dismiss the defect models as candidates for structure formation 
but to find the properties that are responsible for the disagreement and 
try to formulate the constraints that data imposes on the population and 
the properties of topological defects.

As we will illustrate, some of the disagreements 
between the data and the predictions of the cosmic string models can be 
resolved by going to a marginally closed Universe ($\Omega=1.3$) and by 
including the effects of the string decay products and the small-scale 
scale structure on the long strings. These are still 
insufficient to obtain an acceptable fit to the BOOMERANG/MAXIMA data. 
The data can be fit by requiring a higher degree of coherence of 
the string network. None of the currently known string models has the required 
degree of coherence. It would have to be achieved artificially,{\it e.g.}, by
making the coherence time an arbitrary paramter. 
This is not the only way of adding an arbitrary parameter to fit the data. 
In mixed models of strings+inflation \cite{hybrid} the ratio of the
contributions from strings and inflation is an arbitrary parameter and 
in the models with a varying speed of light {\cite{varyingc}) there is
an arbitrariness in the choice of $c_{\phi}$, the speed with which the scalar 
field interacts at the formation of the strings. 
Such models can be rather successful in fitting the 
observations at the cost of violating the causality in its conventional sense.
Namely, in these models the strings are allowed to be correlated on 
superhorizon scales.

Even though these solutions do not have much of an appeal at the moment,
the interest to defect models could be revived should significant non-gaussian
signatures be detected in the CMB.

\section{Predictions of current cosmic string models}
\label{stringpredictions}

The CMB anisotropy from cosmic strings was calculated by a number 
of different research groups\footnote{Considering the large amount of 
papers published
on the subject our list of references cannot possibly be complete. We
refer the reader the recent edition of the book by Vilenkin and Shellard 
\cite{vilshel} for a more comprehensive review of the literature} 
\cite{PenTurokSeljak,aletal,aletal2,hind98,copmagst,allen,uetc,PogVach,steer,avelino}.

The current state of affairs can be inferred by looking at the plots in
Figures \ref{cont1},\ref{cont2},\ref{pogvach1} and \ref{pogvach2}. 
Figures \ref{cont1} and \ref{cont2}
are from \cite{hind98} in which the authors studied the effects of
adding an extra fluid component describing the decay products of the 
long strings. They considered a class of models in which the equation
of state for the extra fluid was parametrized by a single parameter $w^X$.
The choice of $w^X=1/3$ corresponds to the gravitational
radiation. They assumed a flat Friedmann-Roberston-Walker (FRW) universe 
with cold dark matter and no 
cosmological constant. Figures \ref{pogvach1} and \ref{pogvach2} show the 
results of \cite{PogVach} where the wiggly nature of long strings was
taken into account. The plots in Figures \ref{pogvach1} and \ref{pogvach2}
were obtained assuming a flat universe with cold dark matter and 
a cosmological constant ($\Omega_{\Lambda}=0.7$).

\begin{figure}[tbp]
\centerline{\epsfxsize = 0.8\hsize \epsfbox{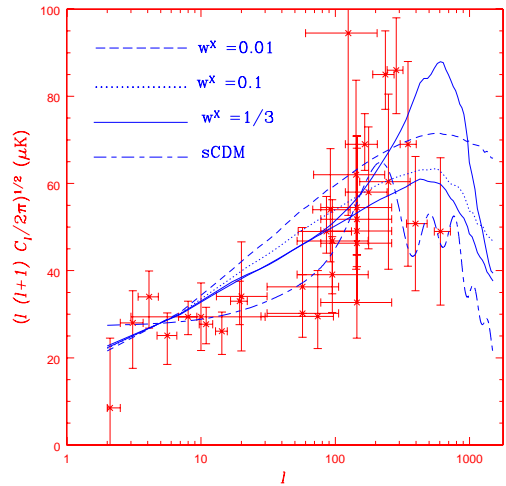}}
\vskip 0.5 truecm
\caption{A plot from \protect \cite{hind98} of the CMB power spectrum for 
cosmic strings.
A flat FRW geometry with no cosmological constant is assumed.
The BOOMERANG and MAXIMA data were not available then.
The higher curve corresponding to $w^X=1/3$ shows what happens if
5\% of the energy goes into the radiation fluid.
} 
\label{cont1}
\end{figure}

\begin{figure}[tbp]
\centerline{\epsfxsize = 0.8\hsize \epsfbox{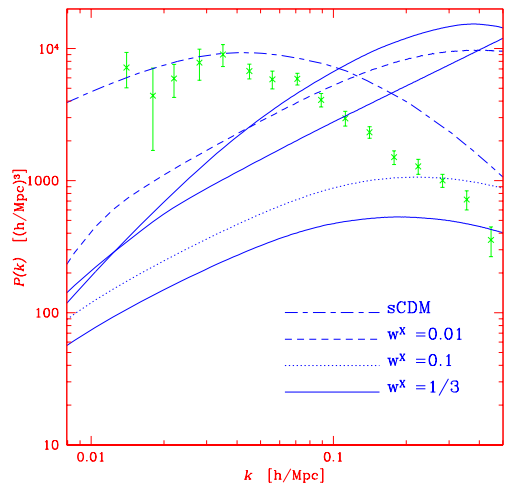}}
\vskip 0.5 truecm
\caption{A plot from \protect \cite{hind98}. The CDM power spectrum
for cosmic strings and the data points inferred by Peacock and Dodds 
from galaxy surveys \protect \cite{pedod}. 
The top 2 $w^X=1/3$ curves correspond to a 5\%
transfer into CDM, and a 20\% transfer into baryons (top).
} 
\label{cont2}
\end{figure}

\begin{figure}[tbp]
\centerline{\epsfxsize = 0.8\hsize \epsfbox{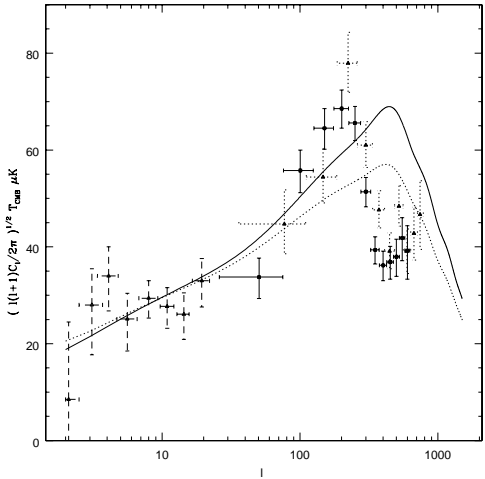}}
\vskip 0.5 truecm
\caption{The CMB power spectrum from \protect \cite{PogVach} 
for smooth (dashed) and wiggly (solid)
cosmic strings with the data from COBE, BOOMERANG and MAXIMA. 
The geometry is flat with $\Omega_{\Lambda}=0.7$.
} 
\label{pogvach1}
\end{figure}

\begin{figure}[tbp]
\centerline{\epsfxsize = 0.8\hsize \epsfbox{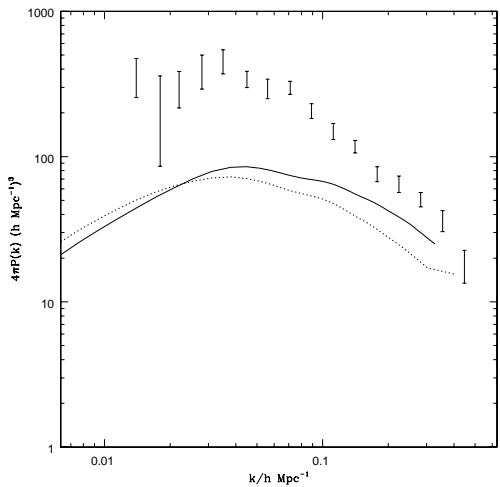}}
\vskip 0.5 truecm
\caption{ The CDM power spectrum from \protect \cite{PogVach}
for smooth (dashed) and wiggly (solid) cosmic strings with the data
from Peacock and Dodds.
} 
\label{pogvach2}
\end{figure}

It is not hard to see that the agreement with the data is poor. The apparent 
problems are:
1) The main peak is too wide and there is no second peak.
2) The main peak in angular spectrum is in the wrong place.
3) The matter power spectrum needs a big bias factor, especially on
large scales. How robust are these features? What can be done in order to improve the 
agreement with the data?

Let us start from the bottom of the list and work our way up.

The solution to the large bias problem in the matter power spectrum 
is partially tied to the solution of the other two problems on the list. 
Certain defect models, in particular, models with a cosmological constant \cite{aletal2} 
\cite{PogVach}, can fit the shape of the data well.  
There is still some uncertainty 
in the data itself as new and more precise measurement are eagerly awaited 
from the SLOAN collaboration \cite{sloan}. Thus, we will 
postpone the detailed analysis of the matter power spectrum predictions.

\section{The structure of Doppler peaks}

The peak structure in the angular spectrum is determined by three main 
factors: the geometry of the universe, coherence and causality.
 
\subsection{Strings in a closed universe}

The curvature of the universe directly affects the paths of light rays 
coming to us from the surface of last scattering \cite{weinberg}. 
In a closed universe,
because of the lensing effect induced by the positive curvature, 
the same physical distances between points on the sky would correspond 
to larger angular scales. As a result, the peak
structure in the CMB angular power spectrum would shift to the 
larger angular scales or the smaller values of $l$.

What are the current observational constraints on the value of $\Omega_{total}$?
Excluding the CMB constraints on the cosmological parameters, 
since they are obtained using the adiabatic inflationary models, 
the current estimates of the matter content of the universe
are (see \cite{parameters} and references therein)
$$
\Omega_{Matter} = 0.35 \pm 0.1 \  \, \ ,
$$
$$
\Omega_{\Lambda} \approx {1\over 3} (4 \Omega_{Matter} + 1) 
\pm 0.15 = 0.8 \pm 0.15.
$$
Adding $\Omega_{Matter}$ and $\Omega_{\Lambda}$ with their
uncertainties gives:
$$
\Omega_{total}\approx 1.15 \pm 0.25 .
$$
Based on this estimate we can conclude that a closed universe with 
$\Omega_{total}=1.3$ is just as probable as a flat one.

\begin{figure}[tbp]
\centerline{\epsfxsize = 0.8\hsize \epsfbox{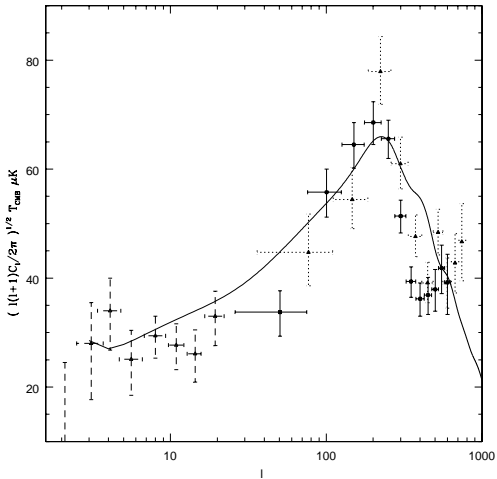}}
\vskip 0.5 truecm
\caption{The CMB power spectrum produced by the wiggly string model of 
\protect \cite{PogVach} in a closed universe with $\Omega_{total}=1.3$. 
$\Omega_{baryon}=0.05$, $\Omega_{CDM}=0.35$, $\Omega_{\Lambda}=0.9$, 
$ H_0 = 65 km s^{-1} Mpc^{-1}$. 
} 
\label{omega}
\end{figure}

In Figure \ref{omega} we plot the prediction of the
cosmic string model of \cite{PogVach} for 
$\Omega_{total}=1.3$.
We used the modified version of CMBFAST \cite{cmbfast} with minor
changes necessary in order to include active sources.
The details of the string model can be found in \cite{PogVach}.
The model was slightly altered by encorporating the recent
improvements to the velocity dependent one-scale model \cite{marshe}
which describes the mean velocity and the correlation length of strings.  
When doing calculations in closed geometry we only included the 
scalar contribution to the angular power spectrum. 
Although the vector and tensor contributions are not negligible
they are sufficiently small for local strings\footnote{The contribution 
of vector modes is much more significant 
in the case of global strings \cite{PenTurokSeljak}.} 
and can be ignored at this point.

As can be seen from Figure \ref{omega} the position of the main peak in 
the angular power spectrum can be matched by choosing a reasonable
value for $\Omega_{total}$. This is in an approximate agreement with
the estimate based on the simple formula for the shift of the peak due
to the curvature effect alone given by
S. Weinberg \cite{weinberg}. This formula, valid for significant 
$\Omega_\Lambda$ and for $\Omega_{total}$ close to $1$, relates the
position of the peak in the flat universe, $l_0$, to the position of the peak
in a curved case, $l_1$, as $l_1=l_0 \Omega^{-1.58}$. There are additional 
important effects which affect the peak structure such as
the changes in the evolution of the string network, which we included in the
simulation.  

Even with the main peak in the right place the agreement with
the data is far from satisfactory. The peak is significantly 
wider than that in the data and there is no sign of a rise in power at
$l\approx 600$. As can be seen from Figs. \ref{cont1} and \ref{pogvach1}, 
the sharpness and the height of the main peak in the angular spectrum
can be enhanced by including the effects of the gravitational radiation 
and wiggles.
More precise high-resolution
numerical simulations of string networks in realistic cosmologies 
with a large
contribution from $\Omega_{\Lambda}$ are needed to determine the exact 
amount of small-scale structure on the strings and the nature of the 
products of their decay. It is, however, 
unlikely that including these effects alone would result in a 
sufficiently narrow main peak and some presence of a second peak. 

This brings us to the issues of causality and coherence and how the random 
nature of the string networks comes into the calculation of the anisotropy 
spectrum.

\subsection{Can cosmic strings play in the orchestra?}

Both, experimental and theoretical results for the CMB power 
spectra involve calculations of averages.
When estimating the correlations of the observed temperature anisotropies 
the experimentalists compute the
average over all available patches on the sky. When calculating the 
predictions of their models theorists find the average over the 
ensemble of possible outcomes provided by the model.   

In inflationary models only the initial conditions for the perturbations 
are random. The subsequent evolution is the same for all members of the 
ensemble. Such models are also called passive. For wavelengths higher 
than the Hubble radius, the linear evolution equations for the Fourier 
components of such perturbations have a growing and a decaying solution. 
The modes corresponding to smaller wavelengths have only oscillating 
solutions.
As a consequence, prior to entering the horizon, each mode undergoes a
period of phase ``squeezing'' which leaves it in a highly coherent state 
by the time it starts to oscillate (see \cite{albrecht} for a nice discussion).
Coherence here means that all members of the ensemble, corresponding 
to the same Fourier mode,
have the same temporal phase. So even though there is randomness
involved, as one has to draw random amplitudes for the oscillations of 
a given mode, the time behavior of different members of the ensemble 
is highly correlated. The total power spectrum is the ensemble averaged
superposition of all Fourier modes and the coherence results in an
interference pattern seen as the Doppler peaks.

In contrast, the evolution of the string network is highly non-linear.
Cosmic strings are expected to move at relativistic speeds, self-intersect 
and reconnect in a chaotic fashion. 
The consequence of this behavior is that the unequal time correlators 
of the string energy-momentum vanish for time differences larger than a
certain coherence time. Members of the ensemble corresponding to a given
mode of perturbations will have random temporal phases with the dice thrown 
on average once in each coherence time. The coherence time of
a realistic string network is rather short.
As a result, the interference pattern in the angular power spectrum is 
completely washed out.

The causality manifests itself, first of all, through the initial conditions 
for the string sources, the perturbations in the metric and the densities of 
different particle species. If one assumes that the defects are formed 
by a causal mechanism in an smooth universe then the correct initial 
condition are obtained by setting the components of the stress-energy 
pseudo-tensor $\tau_{\mu \nu}$ to zero \cite{veersteb}
\cite{PenSperTur}. These are the same as the isocurvature initial conditions 
\cite{HuSperWhite}. 
A generic prediction of isocurvature models (assuming perfect coherence) 
is that the first Doppler is almost completely hidden. 
The main peak is the second Doppler peak and in flat geometries it 
appears at $l\approx 300-400$. This is due to the fact that after entering
the horizon a given Fourier
mode of the source perturbation requires time to induce perturbations in
the photon density. 

The causality also manifests itself through the fact that no superhorizon
correlations in the string energy density are allowed. The correlation
length of a ``realistic'' string network, is normally between 0.1 and 0.4 of
the horizon size.

A interesting study was performed by Magueijo {\it et al} \cite{mag&alb} 
where they have constructed a toy model of defects with two parameters: 
the coherence length and the coherence time. The coherence length was 
taken to be the scale at which the energy density power spectrum of the 
strings turns from a power law decay for large values of $k$ into a white 
noise at low $k$. This is essentially the scale corresponding to the 
correlation length of the string network. The coherence time was defined in 
the sense described in the beginning of this section, in particular, as the 
time difference needed for the unequal time-correlators to vanish. 
\begin{figure}[tbp]
\centerline{\epsfxsize = 1.0\hsize \epsfbox{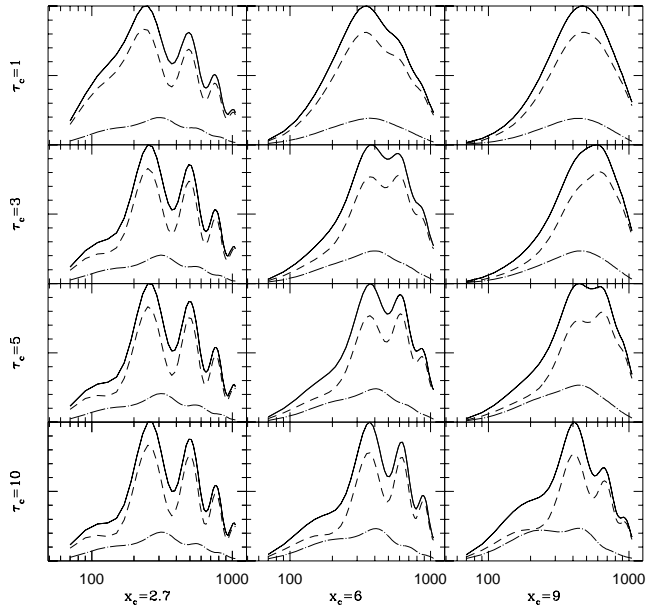}}
\vskip 0.5 truecm
\caption{The predictions of the toy model of Magueijo {\it et al}
\protect \cite{mag&alb} for different values of parameters $x_c$, the coherence
length, and $\tau_c$, the coherence time. $x_c\propto \eta / \lambda_c(\eta)$, 
where $\eta$ is the conformal time and $\lambda_c(\eta)$ is the correlation
length of the network at time $\eta$. One can obtain oscillations in
the CMB power spectrum by fixing either one 
of the parameters and varying the other.
} 
\label{magetal}
\end{figure}
Their study showed (see Figure \ref{magetal}) that by accepting any 
value for one of the parameters and 
varying the other (within the constraints imposed by causality) one can 
reproduce the oscillations in the CMB power spectrum.
Unfortunately for cosmic strings, at least as we know them today, 
they fall into the parameter range corresponding to the upper right corner 
in Figure \ref{magetal}.

In order to fit the observations, the cosmic strings must either be
more coherent or they have to be stretched over larger distances, which
is another way of making them more coherent. To understand this imagine that
there was just one long straight string stretching across the universe
moving with some velocity. The evolution of this string would be linear
and the induced perturbations in the photon density would be coherent.
By increasing the correlation length of the string network we would move
closer to this limiting case of just one long straight string and so the
coherence would be enhanced.

\begin{figure}[tbp]
\centerline{\epsfxsize = 0.8\hsize \epsfbox{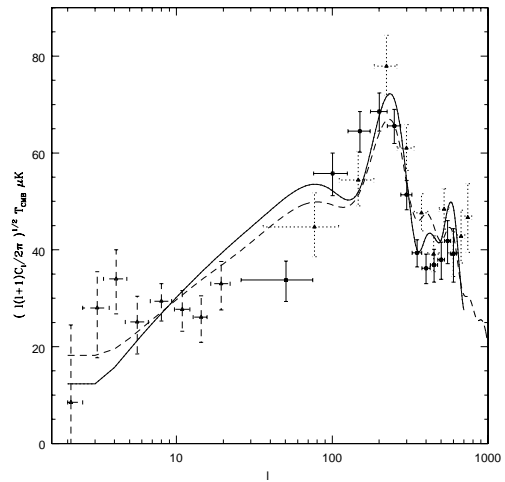}}
\vskip 0.5 truecm
\caption{The CMB power spectrum predicted by a causal cosmic string
model with enhanced coherence properties in a universe with $\Omega=1.3$.
The two different curves correspond to two different values of the
range over which the random phases of the time-oscillations of the
source were allowed to vary. 
} 
\label{fixedphase}

\end{figure}

We have tried experimenting with our string model by artifically making it
more coherent. This can be achieved, for instance, by limiting the range 
over which the random phases of the time oscillations of the source are 
allowed to vary. The results are shown in Fig. \ref{fixedphase}, where we 
have also chosen $\Omega_{total}=1.3$. The agreement with the data is 
somewhat better than that in Figures \ref{cont1}-\ref{omega} but there
is an additional small peak at $l\approx 80$ (the so-called isocurvature peak)
which is not present in the data.
Different string models could, perhaps, produce a closer fit. 
According to the results of \cite{mag&alb}, such models would most 
likely need a larger correlation length in addition to a large coherence
time.

\section{Conclusions}
\label{conclude}

The question of whether the defects can result in a pattern of the CMB power spectrum similar to the Doppler peaks produced by the 
adiabatic inflationary models was repeatedly addressed in the
literature \cite{liddle,turok,mag&alb,varyingc,hybrid}. 
In particular, it was shown \cite{turok,mag&alb} that one can 
construct a causal model of active seeds which
for certain values of parameters can reproduce the oscillations in 
the CMB spectrum.
The problem with the current models of cosmic strings is that they 
fall out of the parameter range that is needed to fit the observations.
 
At the moment, the models with varying speed of light \cite{varyingc} 
and the hybrid models of strings+inflation \cite{hybrid} are the
only models involving topological defects that can match 
the observations. These models require a violation of causality and
contain an arbitrary parameter. Fixing one parameter may 
not sound so terrible
considering the degree of arbitrariness involved in 
inflationary models. But, of course, the defect models would lose 
their most attractive feature, namely, being defined only by the 
symmetry breaking sequence that led to the particle physics we see in 
accelerators.
  
If causality is preserved, then incoherence is the main serious problem faced by the string models. Introducing an arbitrary parameter, 
the coherence time, and assuming a marginally closed universe
may be sufficient to fit the data if the string network has a 
somewhat large correlation length.
Can we think of a credible theory that would result in a more
``time-organized'' string network? Currently, 
we do not know of any physical mechanism that would
make strings more coherent.

In summary, unless some additional new physics is discovered or
postulated, the CMB data excludes the topological defects as the 
primary source of the structure formation.
This does not imply that defects were not involved at all. 
Measurements of the non-gaussian features of the CMB anisotropy 
will further test the predictions of simple inflationary models. 
Certain inflationary models can produce a significant non-gaussian component 
\cite{linde} with a $\chi^2$ distribution. 
A different distribution would mean that a mechanism other 
than inflation took part in seeding the perturbations 
\cite{nongauss1}\cite{nongauss2}.
Future CMB measurements covering a bigger range of angular scales will also tell us if there is really a second and higher order peaks in 
the angular power spectrum. The results from the SLOAN collaboration 
will soon provide us with reliable data for the matter power spectrum. 
With improving computers and modeling techniques more complete high-resolution simulations of the formation and evolution of 
topological defects may point to some new, yet unknown properties.

{\it Acknowledgements:} 
I would like to thank Tanmay Vachaspati for his advice,
J. Magueijo for letting me use the figures from 
\cite{mag&alb} and \cite{hind98},
R. Durrer, M. Sakellariadou, D. Steer, D. Stojkovic, A. Vilenkin and 
S. Winitzki for useful comments.


\begin{thebibliography}{999}

\bibitem{boomerang} P. de Bernardis, {\it et al}, Nature 404, 995 (2000),
astro-ph/0004404

\bibitem{maxima} S. Hanany, {\it et al}, submitted to Ap.J, astro-ph/0005123

\bibitem{jaffe} A.H. Jaffe, {\it et al}, astro-ph/0007333

\bibitem{hybrid} C. Contaldi, M. Hindmarsh, J. Magueijo, Phys.Rev.Lett. {\bf 82}
(1999) 2034-2037; F.R. Bouchet, P. Peter, A. Riazuelo, M. Sakellariadou, 
astro-ph/0005022

\bibitem{albrecht} A. Albrecht, astro-ph/9612015

\bibitem{varyingc} P. P. Avelino, C. J. A. P. Martins, G. Rocha, 
Phys. Lett. {\bf B483} (2000) 210; P.P. Avelino and C.J.A.P. Martins, 
astro-ph/0002413; C.J.A.P Martins, astro-ph/0008287

\bibitem{mag&alb} J. Magueijo, A. Albrecht, P. Ferreira, D. Coulson, 
Phys.Rev. {\bf D54} (1996) 3727-3744

\bibitem{PenSperTur} U.-L. Pen, D. Spergel, N. Turok, Phys. Rev. 
{\bf D49},692 (1994)

\bibitem{HuSperWhite}, W. Hu, D.N. Spergel and M. White,  
Phys.Rev. {\bf D55} (1997) 3288-3302


\bibitem{PenTurokSeljak} N. Turok, U.-L. Pen, U. Seljak, Phys.Rev. {\bf D58}
(1998) 023506

\bibitem{aletal} A. Albrecht, R. Battye, J. Robinson, 
Phys. Rev. Lett., {\bf 79}, 4736 (1997); 
Phys. Rev. {\bf D59}, 023508 (1998).

\bibitem{aletal2} R. Battye, J. Robinson, A. Albrecht,
Phys. Rev. Lett. {\bf 80}, 4847 (1998).

\bibitem{hind98} C. Contaldi, M. Hindmarsh, J. Magueijo, 
Phys. Rev. Lett. {\bf 82}, 679 (1999).

\bibitem{copmagst} E.J. Copeland, J. Magueijo and D.A. Steer,
astro-ph/9903174.

\bibitem{allen} B. Allen, {\it et al}, Phys.Rev.Lett. {\bf 79} (1997) 2624-2627

\bibitem{uetc} U.-L. Pen, U. Seljak, N. Turok, Phys.Rev.Lett.,
 {\bf 79} (1997) 1611-1614

\bibitem{PogVach} L. Pogosian, T. Vachaspati, Phys.Rev. {\bf D60} (1999) 083504


\bibitem{steer} E.J.Copeland, J. Magueijo, D.A. Steer,
 Phys.Rev. {\bf D61} (2000) 063505

\bibitem{avelino}P.P.Avelino, E.P.S.Shellard, J.H.P.Wu, B.Allen,
Phys.Rev. {\bf D60} (1999) 023511

\bibitem{pedod} J.A. Peacock and S.J. Dodds, {\it MNRAS} {\bf 267}, 
1020 (1994).

\bibitem{veersteb} S. Veeraraghavan, A. Stebbins, Ap. J. {\bf 365}, 37, 1990

\bibitem{sloan} D.G. York, {\it et al}. (The SDSS Collaboration), 
astro-ph/0006396, 
To appear in AJ, Sept 2000

\bibitem{weinberg} S. Weinberg, astro-ph/0006276

\bibitem{parameters} M. Turner, astro-ph/9912211; 
J. Primack, astro-ph/0007187

\bibitem{marshe} C.J.A.P. Martins and E.P.S. Shellard, hep-ph/0003298

\bibitem{cmbfast} M. Zaldarriaga, U. Seljak, astro-ph/9911219

\bibitem{turok} N.Turok, Phys.Rev.Lett. {\bf 77}, (1996) 4138-4141; 
Phys.Rev. D54,(1996) 3686-3689 
     
\bibitem{liddle} A. Liddle, Phys.Rev. {\bf D51} (1995) 5347-5351

\bibitem{vilshel} A. Vilenkin, E.P.S. Shellard, ``Cosmic Strings and Other
Topological Defects'', 2nd edition, Cambridge U. Press, 2000    


\bibitem{nongauss1} P. P. Avelino, E. P. S. Shellard, J.H.P.Wu, B. Allen,
          Ap. J. Lett. {\bf 507}, L101 (1998);
           
\bibitem{nongauss2} A. Gangui, L. Pogosian, S. Winitzki, (in preparation)
          
\bibitem{linde} A. Linde, Phys. Lett. {\bf B158}, 375 (1985);
L.Kofman, Phys. Lett. {\bf B173} 400 (1986);
A. Linde and V. Mukhanov, Phys. Rev. {\bf D56}, 535 (1997);
E. Komatsu, D.N. Spergel, astro-ph/0005036, submitted to Phys. Rev. Lett.



\end{thebibliography}
\end{document}